\documentclass[amsmath,amssymb,floatfix,aps,prl,showpacs,preprint]{revtex4}

\usepackage{graphics,graphicx}

\begin{document}

\title{Aggregates of two-dimensional vesicles: Rouleaux and sheets}
\author{P. Ziherl}
\affiliation{Department of Physics, University of Ljubljana, Jadranska 19, SI-1000 Ljubljana, Slovenia\\
Jo\v zef Stefan Institute, Jamova 39, SI-1000 Ljubljana, Slovenia}

\date{\today}

\begin{abstract}
Using both numerical and variational minimization of the bending and adhesion energy of two-dimensional 
lipid vesicles, we study their aggregation, and we find that the stable aggregates include an infinite 
number of vesicles and that they arrange either in a columnar or in a sheet-like structure. We calculate 
the stability diagram and we discuss the modes of transformation between the two types of aggregates, 
showing that they include disintegration as well as intercalation.
\end{abstract}
\maketitle

Since Hooke's {\sl Micrographia}, shapes of aggregates of simple cells have fascinated biologists, 
physicists, and mathematicians alike. In the central part of Thompson's classic {\sl On Growth And 
Form} devoted to the "forms of cells" and "cell-aggregates"~\cite{Thompson17}, the main concept used 
to interpret the striking regularity of many aggregates is surface tension which forces cells to 
minimal-area configurations. Of course, Thompson viewed surface tension merely as a mesoscopic
manifestation of the complex biochemical apparatus and the internal structure of a cell. Similar 
ideas are still being explored: Recently, cells of the developing Drosophila retina were shown to pack 
just like soap bubbles~\cite{Hayashi04}. Another concept related to some biological cells (e.g., the 
red blood cell~\cite{Svetina02}) is that of a vesicle formed by a closed lipid bilayer 
membrane~\cite{Seifert97}. The structure of a vesicle aggregate held together by intermembrane 
attraction is determined by both adhesion and membrane elastic energy rather than by surface energy 
alone as in a soap froth. 

Regardless of their mechanical framework --- soap bubble~\cite{Hayashi04}, lipid vesicle~\cite{Seifert97}, 
or a Potts Hamiltonian~\cite{Glazier93} --- phenomenological models of cell aggregates are most transparent 
within the context of undifferentiated systems with distinct geometry, e.g., early embryonic stages or 
layered tissues of epithelial sheets. Some epithelial tissues consist of a single layer of prismatic 
cells and their salient geometrical feature is their cross-section, essentially a polygonal partition 
of a plane. In many cases, the structural statistics of two-dimensional partitions obey simple empirical 
relationships such as the Lewis law first observed in cucumber epidermis~\cite{Lewis28} and the 
Aboav-Weaire law~\cite{Weaire84}. In a soap-froth-like partition, the link between its structure and the 
local equilibrium is embodied in the Plateau rules, one of the premises used to clarify the above 
laws~\cite{Rivier82,Peshkin91}. On the other hand, no such rules exist for partitions formed by vesicle 
aggregates: Their energy functional is more complicated than that of a soap froth, and their minimal-energy 
configuration is unknown.

In this paper, we study aggregates of two-dimensional vesicles as a model which could elucidate some 
aspects of the in-plane structure of certain simple layered tissues. By employing the theory of elasticity 
combined with the contact-potential intermembrane attraction, we build on existing insight 
into the shape of free two-dimensional vesicles~\cite{Leibler87} and their adhesion on flat 
substrates~\cite{Seifert91}. We focus on vesicles of identical area and perimeter, and we find that the 
stable aggregates consist of infinitely many vesicles. We analyze both columnar and sheet-like infinite 
aggregates, discuss the modes of transformation between them, and construct the stability diagram. 

The model used here is based on the two-di\-men\-si\-onal version of the lipid bilayer bending energy 
$W_b=(K/2)\oint C^2(s)\,{\rm d}s,$ where $K$ is the two-dimensional bending constant and $C$ is the local 
curvature~\cite{Canham70}. The integral is evaluated along the contour of a vesicle subject to two constraints, 
that of a fixed perimeter $L$ and that of a fixed enclosed area $A$. The characteristic length scale is given 
by $R_c$, the radius of a circle of perimeter $L$, such that the reduced vesicle perimeter is normalized to 
$\oint{\rm d}s/2\pi R_c=1$ and the reduced area $a=\int{\rm d}A/\pi R_c^2$ ranges between $0$ and $1$. The 
aggregate energy includes the bending energies of all members and the adhesion energy proportional to the 
total length of the contact lines: $W_a=-\Gamma\sum_i\int_{\rm contact}{\rm d}s_i$~\cite{Seifert91}. Here 
$\Gamma$ is the adhesion strength; the sum runs over all contact lines and the integral over their contours. 
The energy scale is given by $\pi K/R_c$, the bending energy of the reference vesicle with $a=1$, and the 
reduced adhesion strength is $\gamma=\Gamma R_c^2/2K.$

\begin{figure}
\includegraphics{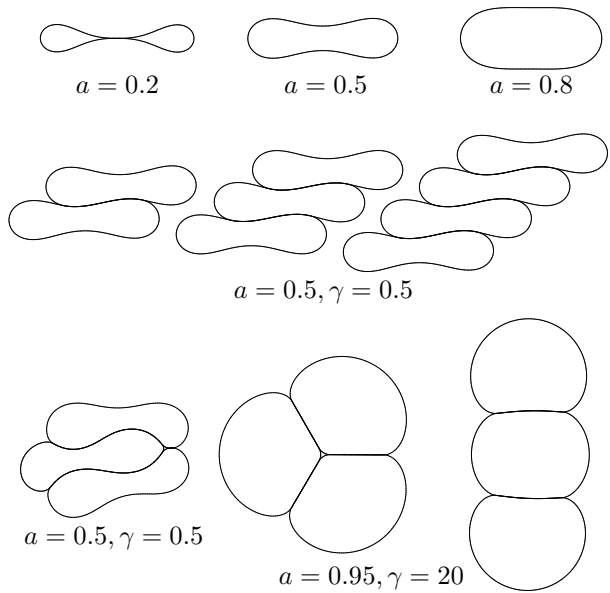}
%\vspace*{80mm}
\caption{Representative numerically obtained vesicle and aggregate shapes. Top: Free vesicles of small ($a=0.2$), 
intermediate ($a=0.5$), and large reduced areas ($a=0.8$). Center: Columnar aggregates of 2, 3, and 4 vesicles 
for $a=0.5$ and reduced adhesion strength $\gamma=0.5$. Bottom: The energy of a trefoil exceeds that of a
columnar triplet unless $a$ and $\gamma$ are both large. The $a=0.5,\gamma=0.5$ trefoil is metastable, its
$a=0.95,\gamma=20$ counterpart is stable. Also shown is the metastable columnar triplet for $a=0.95$ and 
$\gamma=20$.}
\label{shapes}
\end{figure}

By minimizing the energy numerically using Surface Evolver~\cite{Brakke92}, we first reproduce the shapes of 
free vesicles~\cite{Leibler87}. For $a=1$, the vesicle is a circle, for $0.8\lesssim a<1$ it adopts an 
ellipsoidal shape, for $0.27<a\lesssim0.8$ it is characterized by two invaginations on its long sides, and 
for $a<0.27$ it consists of two tether-connected buds (Fig.~\ref{shapes}). Their elongated shape determines 
the way vesicles preferentially stick to each other  --- the contact line of a pair is longest if the 
vesicles' long axes are parallel.

This is confirmed by the doublet shapes; a typical example is shown in Fig.~\ref{shapes}. The doublets show 
that a columnar arrangement (briefly called a {\sl rouleau} due to its equivalence to the characteristic red 
blood cell aggregate) is a good candidate topology. We analyzed its stability as both number of members and 
adhesion strength are varied, finding that the threshold for rouleau formation decreases with the number of 
members. For example, the columnar doublets, triplets, and quadruplets of vesicles with $a=0.5$ are stable 
for reduced adhesion strengths beyond $0.061, 0.052,$ and $0.048$, respectively; the extrapolated aggregation 
threshold for the infinite rouleau is at $\gamma=0.042$. Thus the stable rouleau consists of infinitely many 
vesicles, which is plausible: The more members in an aggregate, the longer the average total contact line per 
vesicle and the lower the threshold.

The rouleau is not the only possible arrangement of vesicles but our numerical studies of other topologies of 
triplets and quadruplets at various $a$ and $\gamma$ strongly indicate that these are relevant exclusively in 
finite aggregates above a certain adhesion strength which increases dramatically as the vesicle area is 
decreased. As an illustration, we note that even in vesicles of area as large as $0.95$, the trefoil is stable 
compared to the rouleau triplet only for $\gamma$ beyond $10.9$ which is well above the aggregation threshold 
$\approx0.5$ as argued below. But the total contact line of the middle vesicle in the $a=0.95$ rouleau triplet 
is longer than the average total contact line per vesicle in the trefoil regardless of $\gamma$, and thus the 
energy of a rouleau with a large enough number of members is always lower than that of a set of trefoils. This 
supports the conclusion that the stable rouleau is infinite.

\begin{figure}[h]
\includegraphics{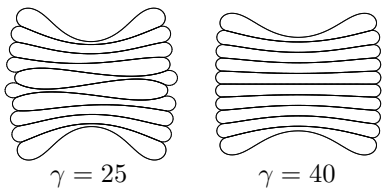}
%\vspace*{30mm}
\caption{Rouleaux of ten $a=0.2$ vesicles for $\gamma=25$ and 40. In both cases, the outer contact lines of the 
central pair are nearly parallel, showing that the pairs can serve as sigmoid-contact and flat-contact repeat 
units of an infinite rouleau.}
\label{ten}
\end{figure}

Fig.~\ref{shapes} shows that the contact lines of vesicles in an infinite rouleau can be either sigmoidal or 
flat~\cite{remark}. This is exemplified by the central pair of vesicles in the numerically obtained ten-member 
$a=0.2$ rouleaux (Fig.~\ref{ten}). The outer contact lines of the pair are almost parallel both at $\gamma=25$ 
where they are sigmoidal and at $\gamma=40$ where they are almost flat, which means that the vesicle pair may 
serve as the repeat unit of an infinite stack. 

\begin{figure}[h]
\includegraphics{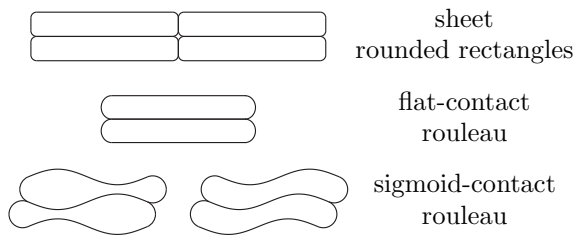}
%\vspace*{40mm}
\caption{Equilibrium $a=0.4$ building blocks of the model sheet-like aggregate (introduced below; $\gamma=45$), 
flat-contact rouleau ($\gamma=19$), and pear-shaped and S-shaped sigmoid-contact rouleau ($\gamma=1$). --- The 
regular tiling shown is a special example of the sheet geometry; sheets are generally structurally disordered as 
the energy of the rounded convex polygonal model vesicle does not depend on its precise shape.}
\label{model}
\end{figure}

Presently, a complete numerical analysis of the structure of vesicle aggregates is well beyond reach even with 
advanced tools such as Surface Evolver. Instead, we use the above cues to construct a variational model of an 
infinite rouleau. We describe the sigmoid-contact vesicle by two circular caps that together make up a full 
circle and two identical one-wave sinusoids combined into a smooth contour (Fig.~\ref{model}), which can be 
done either such that the sinusoids are parallel (which gives an S-shaped vesicle) or such that their phases 
are opposite (which gives a pear-shaped vesicle). The model flat-contact vesicle consists of two identical 
straight lines connected by half-ellipses. --- Using the fixed-area and fixed-perimeter constraints, two out 
of the three parameters of the model shapes can be expressed in terms of the remaining one by resorting to 
analytical approximations of the elliptic integral, and the equilibrium value of the free parameter is then 
adjusted so as to minimize the total energy.

This model reproduces the continuous transition from the sigmoid-contact to the flat-contact repeat unit implied 
by the numerically calculated shapes of the ten-member rouleau for adhesion strengths between 25 and 40. As 
$\gamma$ is increased beyond the transition, the model flat-contact rouleau undergoes a telling transformation: 
The length of the contact line and the eccentricity of the cap of a vesicle both grow such that its shape 
approaches a rounded rectangle (Fig.~\ref{model}). This suggests that at large enough $\gamma$, a sheet-like 
vesicle arrangement with two-dimensional connectivity may be preferred to a rouleau. To explore this possibility, 
we construct a model sheet-like aggregate using convex vesicles with a contour consisting of straight 
contact lines and circular arcs of identical radii (Fig.~\ref{model}). We find that the reduced energy of this 
aggregate is $\sqrt{2\gamma}\left(2-\sqrt{2\gamma}\right)$ independent of $a$, number of sides, and relative 
sizes of the straight sections and the internal angles; the equilibrium curvature of the rounded corners is 
$\sqrt{2\gamma}$ in agreement with the boundary condition at the edge of the contact zone~\cite{Derganc03}. 

\begin{figure}[h]
\includegraphics{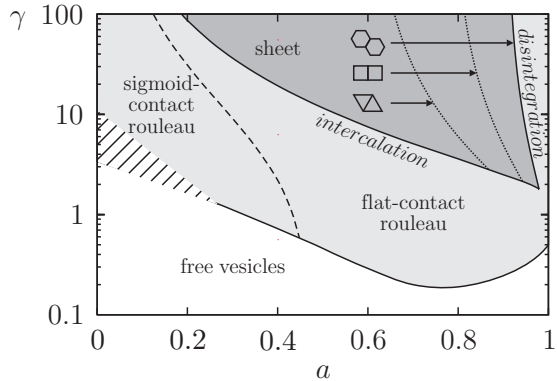}
\caption{Stability map of adhering two-dimensional vesicles is divided into regions of free vesicles, 
infinite rouleaux, and infinite sheet-like aggregates. The sigmoid-contact/flat-contact rouleau transition 
is shown by the dashed line; the large-area limit of stability of rouleaux is infinitesimally smaller than 
$1$ as at $a=1$ the vesicles can only be free. The largest reduced areas where periodic sheet tilings 
with three-, four-, and sixfold coordination are possible correspond to the two dotted lines and the 
right-hand boundary of the sheet region, respectively, labeled by the repeat units free of rounded 
corners for clarity. Also indicated are the two modes of the sheet/rouleau transition described below. 
--- For $a<0.27$, a free vesicle reduces to two tether-connected buds, and in the parameter space 
qualitatively outlined by the crossed region such vesicles form aggregates more complex than those discussed 
here.}
\label{diagram}
\end{figure}

Using all model structures, we can delineate the complete stability diagram of the aggregates; by comparing 
them to the numerically calculated free vesicle shapes, we also outline the aggregation threshold. As 
anticipated, the sequence observed upon increasing adhesion strength is free vesicle/rouleau/sheet unless the 
reduced vesicle area is too small or too large (Fig.~\ref{diagram}); for $a<0.45$, the rouleau region is 
subdivided into the sigmoid-contact and the flat-contact part. For $a<0.27$ where the free vesicles consist 
of two tether-connected buds, the small-$\gamma$ rouleaux are more complicated than those studied here; we did 
not explore this regime in detail.

Given that the rounded polygonal model vesicles are energy degenerate, the sheet-like aggregate should consist 
of coexisting triangles, tetragons, pentagons, hexagons, heptagons, etc. conceivably forming orientationally 
ordered structures. Such coexistence is characteristic of the most part of the sheet region. However, 
as $a$ increases the set of possible shapes is gradually narrowed and the sheet becomes more ordered, which 
is illustrated by the maximal vesicle areas where a periodic tiling with three-, four-, and sixfold coordination 
is possible at a given $\gamma$ (Fig.~\ref{diagram}). The largest reduced vesicle area which supports a sheet 
corresponds to a rounded regular hexagon: Beyond this point, vesicles can only form flat-contact rouleaux. 

A potentially relevant feature of a sheet-like aggregate is the relative intervesicular area $I$ defined as 
the fraction of the plane not covered by the vesicles. $I$ can be evaluated easily for regular tilings: 
To lowest order, $I=\beta/\gamma a$ where $\beta$ equals $(2\sqrt{3}-\pi)/2\pi\approx0.051$ for a tiling of 
equiangular convex hexagons and $(4-\pi)/2\pi\approx0.137$ for a rectangular tiling (suggesting that if there 
were a reason that $I$ should be minimized, partitions with regular 3-way vertices would be preferred). $I$ is 
typically small --- for equiangular hexagons, its upper limit reached at $a=0.98$ and $\gamma=1.8$ is $I=2.9\%$.

On a quantitative note, the aggregation threshold is overestimated in our analysis. For example, the numerically 
obtained value for vesicles with $a=0.5$ is at $\gamma=0.042$ rather than at $\gamma=0.46$ as shown in 
Fig.~\ref{diagram}. This discrepancy is expected as the variational aggregate energies are too large; we 
estimate that for $\gamma\gtrsim0.5$ they depart from the exact values by less than 10\%. However, we stress 
that the error is a rapidly decreasing function of $\gamma$. This is illustrated by the finite rouleaux with 
$\gamma=25$ and $40$ shown in Fig.~\ref{ten} where the central vesicles are clearly described rather well by 
the ansatz shapes compared to vesicles in the $\gamma=0.5$ linear aggregates (Fig.~\ref{shapes}); the 
numerically calculated transition between the two types of the 10-member rouleau with $a=0.2$ is at 
$\gamma\approx32$ very close to $\gamma=31$ as predicted variationally for the infinite rouleau. Thus the model 
performs well at large $\gamma$ and we trust that the sheet/rouleau boundary shown in Fig.~\ref{diagram} is 
quite accurate. 

The sheet/rouleau transition is a very interesting aspect of the stability diagram, and it can be accomplished 
either by disintegration of a sheet into several rouleaux or by vesicle intercalation which transforms a sheet 
into a single rouleau. The first mode is preferred along the right-hand boundary of the sheet region where the 
only possible vesicle type is the rounded hexagon and the sheet can break up into parallel rouleaux along a 
lattice axis. On the other hand, at the left-hand/bottom boundary the sheet is disordered so the vesicles cannot 
dissolve easily into separate rouleaux. But they can intercalate, thereby gradually decreasing the extent of a 
sheet along the average long axis of vesicles. Intercalation is accompanied by vesicle alignment and effectively 
continuous: Relocation of vesicles is facilitated by the degeneracy of their energy so that they may change 
their shape at fixed area and perimeter as appropriate at no cost. 

These results may be related to an important mechanism of cell rearrangement known as convergent 
extension~\cite{Keller02} whereby cells in a layered structure i) undergo in-plane elongation, ii) align, 
and iii) intercalate such that the tissue is extended in the direction perpendicular to the long axes of cells. 
The physical side of this process can be described in terms of the differential adhesion hypothesis by assuming 
that the en-face cross-section of cells is elongated, say rectangular with 2 long and 2 short sides, and that the 
adhesion strengths for long-long, long-short, and short-short contacts are all different~\cite{Zajac00}. 
Under certain conditions, this gives a minimal-energy aggregate that is orientationally ordered and elongated 
normal to cell axes, thus reproducing the final stage of convergent extension. But the simplest microscopic 
basis of differential adhesion would be a variation of the density of anchoring junctions across the cell sides, 
and there is no solid evidence either for or against this~\cite{Zajac00}. 

The stability diagram (Fig.~\ref{diagram}) offers an alternative explanation: Mechanically, convergent extension 
could be the process of sheet/rouleau transition induced by a decrease of either adhesion strength or reduced area 
of the en-face cell cross-section, whereby the sheet would seek the energy minimum by transforming into a rouleau 
via intercalation. In real tissues, convergent extension is certainly not driven solely by the interplay of the 
intermembrane adhesion and cell elastic energy as a passive, non-specific morphogenetic force; the sheet/rouleau 
transition may be either promoted or impeded by the action of cytoskeletal machinery and cell internal structures 
as well as by the interaction of the cell with the extracellular matrix and the surrounding tissues. Nonetheless, 
our scenario is consistent with several subdominant features of the process unaccounted for by the differential 
adhesion model. For example, convergent extension can occur in absence of changes in the cell shape~\cite{Irvine94} 
(in this case, it could be induced by a decrease of adhesion strength), and it can be accompanied by the onset 
of disorder where six-fold cell coordination is replaced by patterns including a range of polygonal 
shapes~\cite{Zallen04} (which happens in the sheet-like aggregate as the vesicle reduced area is decreased). 
This suggests that the differential adhesion model~\cite{Zajac00} can be regarded as an effective, 
coarse-grained version of our theory where the vesicle elastic energy is absorbed in an anisotropic adhesion 
constant.

In conclusion, our analysis of adhering two-dimensional vesicles shows that the structure of minimal-energy 
aggregate is simplified by the fact that the stable aggregates, both linear and sheet-like, are infinite. The 
main results should be qualitatively insensitive to a moderate vesicle polydispersity, and they call for a 
generalization to three-dimensional vesicles. 

\acknowledgments
 
It is a pleasure to thank S. Svetina for a number of stimulating and productive discussions as well as for critical 
reading of the manuscript. This work was supported by Slovenian Research Agency through Grant P1-0055.


\begin{thebibliography}{30}
\bibitem{remark} We tacitly assume that any two vesicles touch at a single contact line, which is true at 
large enough reduced areas. For $a<0.27$ where a free vesicle consists of two tether-connected buds, this need 
not be the case: At small adhesion strengths, vesicles in an aggregate are not too different from the 
free ones, and each bud should form a separate contact with another vesicle. But at large $\gamma$, the vesicles 
are flattened out and the aggregates are qualitatively the same as for $a>0.27$, which is shown in Fig.~\ref{ten}.
\end{thebibliography}
\end{document}